\documentclass[reprint,amsmath,amssymb,aps,floatfix,showkeys,showpacs,
]{revtex4-1}

\usepackage[T1]{fontenc}
\usepackage[utf8]{inputenc}

\usepackage{graphicx,colordvi}
\usepackage{dcolumn}
\usepackage{bm}

\begin{document}

\title{ Finite element method for extended KdV equations}

\author{Anna Karczewska}
 \email{A.Karczewska@wmie.uz.zgora.pl}
\affiliation{Faculty of Mathematics, Computer Science and Econometrics\\ University of Zielona G\'ora, Szafrana 4a, 65-246 Zielona G\'ora, Poland}

\author{Piotr Rozmej}
 \email{P.Rozmej@if.uz.zgora.pl}
\affiliation{Institute of Physics, Faculty of Physics and Astronomy \\
University of Zielona G\'ora, Szafrana 4a, 65-246 Zielona G\'ora, Poland}
\date{\today} 

\author{Maciej  Szczeci\'nski} \email{m.szczecinski@wmie.uz.zgora.pl}
\affiliation{Faculty of Mathematics, Computer Science and Econometrics\\ University of Zielona G\'ora, Szafrana 4a, 65-246 Zielona G\'ora, Poland}

\author{Bartosz Boguniewicz}
 \email{b.boguniewicz@gmail.com}
\affiliation{Institute of Physics, Faculty of Physics and Astronomy \\
University of Zielona G\'ora, Szafrana 4a, 65-246 Zielona G\'ora, Poland}

%\date{\today} 

\begin{abstract}
The finite element method (FEM) is applied to obtain numerical solutions to a recently derived nonlinear equation for the shallow water wave problem. A weak formulation and the Petrov-Galerkin method are used. It is shown that the FEM gives a reasonable description of the wave dynamics of soliton waves governed by extended KdV equations. Some new results for several cases of bottom shapes are presented.
The numerical scheme presented here is suitable for taking into account stochastic effects, which will be discussed in a subsequent paper.  
\end{abstract}


\maketitle

\section{Introduction}\label{Int}

The Korteveg--de Vries equation appears as a model for the propagation of weakly nonlinear dispersive waves in several fields. Among them there are gravity driven waves on a surface of  an incompressible irrotational inviscid fluid \cite{Whit,EIGR,Rem,DrJ,MS90,Ding}, ion acoustic waves in plasma  \cite{EIGR}, impulse propagation in electric circuits \cite{Rem} and so on. 
In the shallow water wave problem the KdV equation corresponds to the case when the bottom is even. % water depth is constant. 
There have been many attempts to study nonlinear waves in the case of an uneven bottom because of its significance, for instance in such phenomena as tsunamis.
Among the first papers dealing with a slowly varying bottom are papers of Mei and Le M\'ehaut\'e \cite{Mei} and Grimshaw \cite{Grim70}. When taking an appropriate average of vertical variables one arrives at Green-Nagdi type equations \cite{GN,Nad,Kim}.
Van Groesen and Pudyaprasetya \cite{G&P1,G&P2} studied uni-directional waves over a slowly varying bottom within the Hamilton approach, obtaining a forced KdV-type equation. An extensive study of wave propagation over an uneven bottom conducted before 2000 is summarized in Dingemans's monograph \cite{Ding}.
The papers \cite{Pel,Peli,Peli1} are examples of approaches that combine linear and nonlinear theories. The Gardner equation and the forced KdV equation, were also extensively investigated in this context, see, e.g., \ \cite{Grim,Smy,Kam}.

In  previous papers, \cite{KRR,KRI} we derived a new KdV-type equation containing  terms which come directly from an uneven bottom. These terms, however, appear naturally only if Euler equations for the fluid motion are considered up to second order in small parameters, whereas the KdV equation is obtained in first order approximation.   
There are no analytic solutions for the above equation. In \cite{KRR,KRI} we presented several cases of numerical simulations for that equation obtained using the finite difference method (FDM) with periodic boundary conditions.

It was demonstrated in \cite{DebP} that finite element method (FEM) describes properly the dynamics of the KdV equation 
%in standard, mathematical form  
(\ref{kdvm}), which is the equation in a moving frame of reference. 

The first aim of this paper is to construct an effective FEM method for solving higher order KdV equations, both with even bottom %(\ref{etaab}) 
and uneven bottom. %(\ref{etaabd})
The second goal is to compare the results obtained in this numerical scheme with some of the results obtained earlier using the finite difference method in \cite{KRR} and in \cite{KRI}.

The paper is organized as follows:  In section \ref{prel} we review the KdV equation (\ref{kdv1}), the extended KdV equation (\ref{etaab}) and KdV-type equation containing direct terms from bottom variation (\ref{etaabd}), all expressed in scaled dimensionless variables. In section \ref{numM} the construction of the numerical method for solving these equations within the FEM is described. Coupled sets of nonlinear equations for coefficients of expansion of solutions to these equations in a basis of piecewise linear functions are obtained. In section \ref{nsym} several examples of numerical simulations are presented.

\section{Preliminaries}\label{prel}

Extended KdV type equations, derived by some of the authors in
 \cite{KRR,KRI}, second order in small parameters, have the following form 
(written in scaled dimensionless coordinates, in a fixed coordinate system).
For the case with an uneven bottom
\begin{eqnarray} \label{etaabd}
\eta_t &+&
\eta_x + \alpha\, \frac{3}{2}\eta\eta_x +\beta\,\frac{1}{6} \eta_{3x} \\
&-&
\frac{3}{8} \alpha^2 \eta^2\eta_x 
+ \alpha\beta \!\left(\!\frac{23}{24}\eta_x\eta_{2x}\!+\!\frac{5}{12}\eta\eta_{3x}\! \right)\!+\!\frac{19}{360}\beta^2\eta_{5x} \nonumber \\
&+&
\beta\delta\left(-\frac{1}{2\beta}(h\eta)_x +\frac{1}{4} \left(h_{2x}\eta\right)_x -\frac{1}{4} \left(h\eta_{2x}\right)_x\right) =0.
\nonumber
\end{eqnarray}
Details of the derivation of the second order equation (\ref{etaabd}) from the set of Euler equations with appropriate boundary conditions can be found in \cite{KRR,KRI}.
In (\ref{etaabd}), $\eta(x,t)$ stands for a wave profile and $h=h(x)$ denotes a bottom profile. Subscripts are used for notation of partial derivatives, that is, for instance $\eta_{2x}\equiv \frac{\partial^2 \eta}{\partial x^2}$, and so on.
Small parameters ~$\alpha,\beta,\delta$ are defined by ratios of the amplitude of the wave profile ~$a$, the depth of undisturbed water  ~$h_0$, average wavelength ~$l$  and the amplitude of the bottom changes ~$a_h$
\begin{equation} \label{smallp}
 \alpha =\frac{a}{h_0}, \qquad \beta=\left(\frac{h}{l}\right)^2, \qquad \delta=\frac{a_h}{h_0}.
\end{equation}
For details of the transformation of the original dimensional variables to the nondimensional, scaled ones used here, see, e.g.,\ \cite{KRR,KRI,BS}.

It should be emphasized that in equation (\ref{etaabd}) all three terms originating from an uneven bottom are second order in small parameters. These terms appear from the boundary condition at the bottom which is already in second order with coefficient $\beta\delta$, see
equation (5) in \cite{KRI} or equation (10) in \cite{KRR}.
Then in  the final second order equation (\ref{etaabd}) we write them in the form $\beta\delta (\cdot)$ in order to epmhasize that they all come from the second order perturbation approach. For details we refer to the mentioned papers. 

In the case of an even bottom ($\delta=0$)  equation (\ref{etaabd}) is reduced to the second order KdV type equation 
\begin{eqnarray} \label{etaab}
\eta_t &+& \eta_x + \alpha\, \frac{3}{2}\eta\eta_x 
+\beta\,\frac{1}{6} \eta_{3x} - \frac{3}{8} \alpha^2 \eta^2\eta_x\\
\! \!&\! \!+ & 
 \alpha\beta\left(\frac{23}{24}\eta_x\eta_{2x}+\frac{5}{12}\eta\eta_{3x} \right)+\frac{19}{360}\beta^2\eta_{5x} = 0 \nonumber 
\end{eqnarray}
and when $\beta=\alpha$ it becomes identical to  Eq.~(21) in \cite{BS}.
Equation (\ref{etaab}) was obtained earlier by Marchant and Smyth  \cite{MS90} and called the {\em extended KdV} equation.

Limitation to first order approximation in small parameters gives the KdV  in a fixed system of coordinates 
\begin{equation} \label{kdv1}
\eta_t+\eta_x + \alpha\, \frac{3}{2}\eta\eta_x +\beta\,\frac{1}{6} \eta_{3x} =0.
\end{equation}

The standard, mathematical form of the KdV equation is obtained from (\ref{kdv1}) by transformation to a moving reference frame. Substituting
\begin{equation} \label{tr}
\bar{x} =\sqrt{\frac{3}{2}}(x-t), \qquad \bar{t}=\frac{1}{4}\sqrt{\frac{3}{2}}\,\alpha\,t, \qquad u = \eta,
\end{equation}
one obtains from (\ref{etaab}) the equation
\begin{equation} \label{kdvm}
u_{\bar{t}} + 6\,u\,u_{\bar{x}} + \frac{\beta}{\alpha}u_{3\bar{x}}%\bar{x}\bar{x}}
 =0,
\end{equation}
or finally,  when  ~$\beta=\alpha$,
\begin{equation} \label{kdvm1}
 u_{\bar{t}} + 6\,u\,u_{\bar{x}} + u_{3\bar{x}}=0. %\bar{x}\bar{x}} =0.
\end{equation}
%in {\sl moving frame}.

In this paper we attempt to solve numerically the equation (\ref{etaabd}) for several cases of bottom topography and different initial conditions. 
In several points we follow the method applied by Debussche and Printems~\cite{DebP}. However, the method is extended to higher order KdV type equations with  plain bottom (\ref{etaab}) and with bottom fluctuations  (\ref{etaabd}).
For both cases we work in a fixed reference system, necessary for a bottom profile depending on the position.

\section{Numerical method} \label{numM}

The emergence of soliton solutions to the KdV equation was observed in numerics fifty years ago by \cite{ZK}. Several numerical methods used for solving the KdV equation are  discussed in \cite{TaAb}. Among them are the finite difference explicit method \cite{ZK}, the finite difference implicit method \cite{Goda} and several versions of the pseudospectral method, as in \cite{FoWhi}.
It is also worth mentioning papers using the FEM and Galerkin methods \cite{BoCh,CuMa}. Most numerical applications use periodic boundary conditions, but
there exist also works that apply Dirichlet boundary conditions on a finite interval \cite{SkKa,YiHu,YuSh}.

The authors are trying to construct a method which will be applicable not only for the numerical simulation of an evolution of nonlinear waves governed by equations (\ref{etaabd}) or   (\ref{etaab}) but also for their stochastic versions. Such stochastic equations will be studied in the next paper. %\cite{KRSB2}.  
Since stochastic noise is irregular, solutions are not necessarily smooth, neither in time nor space. A finite element method (FEM) seems to be suitable for such a case. 

\subsection{Time discretization}  \label{timd} 
We have adapted the Crank--\-Nicholson scheme for time evolution, beginning with the KdV equation (\ref{kdv1}) in a fixed coordinate system. Note that $\eta\eta_x=\frac{1}{2}(\eta^2)_x$. Denote also $v:=\eta_x$ and $w:=v_x$. Let us choose time step $\tau$. Then the KdV equation (\ref{kdv1}) in the Crank--Nicholson scheme  can be written as a set of coupled first order differential equations
\begin{eqnarray} \label{Cr}
\eta^{n+1}- \eta^{n}+\tau  \left(\frac{\partial }{\partial x}\eta^{n+\frac{1}{2}} \right.\hspace{10ex}\nonumber\\ \left. \hspace{2ex}
+ \frac{3\alpha}{4} 
\frac{\partial }{\partial x}(\eta^{n + \frac{1}{2}})^{2} +  \frac{\beta}{6} w^{n + \frac{1}{2}} \right) &=& 0,  \\ %\label{Cr1}
\frac{\partial }{\partial x}\eta^{n + \frac{1}{2}} &=& v^{n + \frac{1}{2}}, \hspace{4ex} \nonumber\\ %\label{Cr2}
\frac{\partial }{\partial x}v^{n + \frac{1}{2}} &=&  w^{n + \frac{1}{2}},  \hspace{4ex}\nonumber
\end{eqnarray}
where 
\begin{equation} \label{e12a}\def\arraystretch{1.4} 
\begin{array}{rcl}
\eta^{n+\frac{1}{2}}&=&\frac{1}{2}\left( \eta^{n+1} + \eta^{n} \right),  \\
v^{n+\frac{1}{2}}&=&\frac{1}{2}\left( v^{n+1} + v^{n} \right),  \\
w^{n+\frac{1}{2}}&=&\frac{1}{2}\left( w^{n+1} + w^{n} \right). 
\end{array}
\end{equation}

For second order equations (\ref{etaabd}) or (\ref{etaab}) we need to introduce two new auxiliary variables: $p:=w_x$ and $q:=p_x$. 
Note that $\eta^2\eta_x=\frac{1}{3}(\eta^3)_x$, ~$\eta_x\eta_{2x}=\frac{1}{2}(\eta_x^2)_x= \frac{1}{2}(v^2)_x$.  Moreover, $\eta_{5x}=q=p_x$ and
$$\frac{23}{24}\eta_x\eta_{2x}+\frac{5}{12}\eta\eta_{3x}=\frac{13}{48}(v^2)_x+\frac{5}{12}(\eta w)_x.$$

This setting allows us to write the Crank--\-Nicholson scheme for (\ref{etaab}) as the following set of first order equations
\begin{eqnarray} \label{CrAB}
\eta^{n+1}- \eta^{n}+ \hspace{28ex}\nonumber \\
+\tau \frac{\partial }{\partial x} \left[ \eta^{n+\frac{1}{2}} +\frac{3\alpha}{4} 
\left(\eta^{n + \frac{1}{2}}\right)^{2} 
 + \! \frac{\beta}{6} w^{n + \frac{1}{2}} \!  \right. \nonumber \\ \hspace{-3ex}
-\!\frac{1}{8}\alpha^2 \left(\eta^{n+\frac{1}{2}}\right)^3
+ \alpha\beta\left(\!\frac{13}{48}\left(v^{n + \frac{1}{2}}\right)^2  \right. \\ \left. \left.\hspace{-3ex}
\!+\!\frac{5}{12}\left(\eta^{n + \frac{1}{2}}w^{n + \frac{1}{2}\!}\right)\! \right)\! 
+ \frac{19}{360}\beta^2 \left(q^{n + \frac{1}{2}}\right)
 \right]
&=& 0, \nonumber\\
\frac{\partial }{\partial x}\eta^{n + \frac{1}{2}}-v^{n + \frac{1}{2}} &=& 0,  \nonumber\\
\frac{\partial }{\partial x} v^{n + \frac{1}{2}}-w^{n + \frac{1}{2}} &=&  0,  \nonumber \\
\frac{\partial }{\partial x} w^{n + \frac{1}{2}}-p^{n + \frac{1}{2}} &=& 0,  \nonumber\\ \frac{\partial }{\partial x} p^{n + \frac{1}{2}}-q^{n + \frac{1}{2}} &=& 0,  \nonumber 
\end{eqnarray}
where
%\begin{eqnarray} \label{e12}p^{n+\frac{1}{2}}&=&\frac{1}{2}\left( p^{n+1} + p^{n} \right),  \\q^{n+\frac{1}{2}}   &=&\frac{1}{2}\left( q^{n+1} + q^{n} \right). \nonumber \end{eqnarray}
\begin{equation} \label{e12}\def\arraystretch{1.4} 
\begin{array}{rcl}
p^{n+\frac{1}{2}}&=&\frac{1}{2}\left( p^{n+1} + p^{n} \right),  \\
q^{n+\frac{1}{2}}   &=&\frac{1}{2}\left( q^{n+1} + q^{n} \right). 
\end{array}
\end{equation}

For the second order KdV type equation with an uneven bottom (\ref{etaabd}) the first equation in the set (\ref{CrAB}) has to be supplemented by terms originating from bottom variations, yielding
\begin{eqnarray} \label{CrABD}
\eta^{n+1}- \eta^{n}+ \hspace{28ex}\nonumber \\
+\tau \frac{\partial }{\partial x} \left[ \eta^{n+\frac{1}{2}} +\frac{3\alpha}{4} 
\left(\eta^{n + \frac{1}{2}}\right)^{2} 
 + \! \frac{\beta}{6} w^{n + \frac{1}{2}} \!  \right. \nonumber \\ \hspace{-3ex}
-\!\frac{1}{8}\alpha^2 \left(\eta^{n+\frac{1}{2}}\right)^3
+ \alpha\beta\left(\!\frac{13}{48}\left(v^{n + \frac{1}{2}}\right)^2  \right. 
\\ \left. \hspace{-3ex}
\!+\!\frac{5}{12}\left(\eta^{n + \frac{1}{2}}w^{n + \frac{1}{2}\!}\right)\! \right)\! 
+ \frac{19}{360}\beta^2 \left(q^{n + \frac{1}{2}}\right)  \nonumber 
\\  
\frac{1}{4}\beta\delta  \left(-\frac{2}{\beta}\left(h^{n + \frac{1}{2}}\eta^{n + \frac{1}{2}\!}\right)  \right. \hspace{5ex} \nonumber \\ \left. \left.
+\eta^{n + \frac{1}{2}}g^{n + \frac{1}{2}} + h^{n + \frac{1}{2}}w^{n + \frac{1}{2}} \right) \right]
&=& 0, \nonumber
\end{eqnarray} 
where ~$g:=h_{xx}$.

Below we focus on the second order equations (\ref{etaab}) and (\ref{CrAB}), pointing out contributions from bottom variation later.

\subsection{Space discretization}  \label{spad}
Following the arguments given by Debussche and Printems \cite{DebP} we apply the Petrov-Galerkin discretization and finite element method. We use  piecewise linear shape functions and piecewise constant test functions. We consider wave motion on the interval $x\in [0,L]$ with periodic boundary conditions. Given $N\in \mathbb{N}$, then we use a mesh $M_{\chi}$ of points $x_j= j\chi$, $j=0,1,\ldots,N$, where $\chi =L/N$. Let $V^1_{\chi}$ which is a space of  piecewise linear functions $\varphi_j(x)$, such that $\varphi_j(0)=\varphi_j(L)$, defined as
\begin{equation} \label{phi}
\varphi_{j}(x) = \left\{ \begin{array}{lll}
\frac{1}{\chi}(x-x_{j-1}) & \mbox{if} &  x \in [x_{j-1},x_{j}] \\
\frac{1}{\chi}(x_{j+1}-x) & \mbox{if} &  x \in [x_{j},x_{j+1}] \\
0 &  & \mbox{otherwise} . \end{array} \right. 
\end{equation}
As test functions we have chosen the space of piecewise constant functions $\psi_j(x)\in V^0_{\chi}$, where 
\begin{equation} \label{psi}
\psi_{j}(x) = \left\{ \begin{array}{lll}
1 & \mbox{if} &  x \in [x_{j},x_{j+1})\\
0 &  &\mbox{otherwise}. \end{array} \right.
\end{equation}

An approximate solution and its derivatives may be written as an expansion in the basis  (\ref{phi}) 
\begin{equation} \label{etfi}\def\arraystretch{1.4} 
\begin{array}{lcl}
\eta_{\chi}^{n}(x)& = &\sum_{j=1}^{N}a_j^n\,\varphi_j(x),\\
 v_{\chi}^{n}(x)& = &\sum_{j=1}^{N}b_j^n\,\varphi_j(x),\\
 w_{\chi}^{n}(x)& = &\sum_{j=1}^{N}c_j^n\,\varphi_j(x),\\
 p_{\chi}^{n}(x)& = &\sum_{j=1}^{N}d_j^n\,\varphi_j(x),\\
 q_{\chi}^{n}(x)& = &\sum_{j=1}^{N}e_j^n\,\varphi_j(x),
\end{array} 
\end{equation}
where $a^n_j,b^n_j ,c^n_j ,d^n_j , e^n_j$ are expansion coefficients.
Therefore, in a weak formulation we can write (\ref{Cr}) as
\begin{eqnarray} \label{CrABw}
\left(\eta_{\chi}^{n+1}- \eta_{\chi}^{n},\psi_i\right)+ \tau \left\{   \left(\partial_x\eta_{\chi}^{n+\frac{1}{2}},\psi_i\right)  \right. \hspace{10ex} \nonumber \\
  +\frac{3\alpha}{4} \left(
\partial_x\left(\eta_{\chi}^{n + \frac{1}{2}}\right)^{2},\psi_i\right) 
 +  \frac{\beta}{6}\left(\partial_x w_{\chi}^{n + \frac{1}{2}},\psi_i\right)    \nonumber \\ \hspace{-3ex}
-\frac{1}{8}\alpha^2 \left(\partial_x\left(\eta_{\chi}^{n+\frac{1}{2}}\right)^3,\psi_i\right)\nonumber  \\
+ \alpha\beta\left[\frac{13}{48}\left(\partial_x\left(v_{\chi}^{n + \frac{1}{2}}\right)^2,\psi_i\right)  \right. \\  \left.\hspace{-3ex}
+\frac{5}{12}\left(\partial_x\left(\eta_{\chi}^{n + \frac{1}{2}}w_{\chi}^{n + \frac{1}{2}}\right),\psi_i\right)    \right] \nonumber  \\ \left.
+ \frac{19}{360}\beta^2\left( \partial_x\left(q_{\chi}^{n + \frac{1}{2}}\right),\psi_i\right) \right\}
 &=&      0, \nonumber\\
\left(\partial_x\eta_{\chi}^{n + \frac{1}{2}},\psi_i\right)-\left(\!v_{\chi}^{n + \frac{1}{2}},\psi_i\!\right) &=& 0, \nonumber\\ 
\left(\partial_x v_{\chi}^{n + \frac{1}{2}},\psi_i\right)-\left(\!w_{\chi}^{n + \frac{1}{2}},\psi_i\!\right) &=&0, \nonumber \\
\left(\partial_x w_{\chi}^{n + \frac{1}{2}},\psi_i\right)-\left(\!p_{\chi}^{n + \frac{1}{2}},\psi_i\!\right) &=& 0,  \nonumber\\ 
\left(\partial_x p_{\chi}^{n + \frac{1}{2}},\psi_i\right)-\left(\!q_{\chi}^{n + \frac{1}{2}},\psi_i\!\right)  &=& 0, \nonumber 
\end{eqnarray}
for any $i=1,\dots,N$, where for abbreviation  $\partial_x$ is used for $\frac{\partial }{\partial x}$.
In (\ref{CrABw}) and below scalar products are defined by appropriate integrals
$$(f,g):=\int_0^L f(x) g(x) dx.$$ 

In the case of 
equation (\ref{etaabd}), the first equation of the set (\ref{CrABw}) has to be supplemented  inside the bracket \{  \}    by the terms
\begin{eqnarray} \label{CrABdno}
 &+& \frac{1}{4}\beta\delta  \left(\partial_x\left[-\frac{2}{\beta}\left(h^{n + \frac{1}{2}}\eta_{\chi}^{n + \frac{1}{2}\!}\right)  \right.\right. \\
&&\hspace{9ex} \left.\left. +\eta_{\chi}^{n + \frac{1}{2}}g^{n + \frac{1}{2}} + h^{n + \frac{1}{2}}w_{\chi}^{n + \frac{1}{2}} \right],\psi_i\right) . \nonumber
\end{eqnarray}
%inside the bracket \{  \}.

Insertion of (\ref{etfi}) into (\ref{CrABw}) yields a system of coupled linear equations for coefficients $a_j^n, b_j^n,  c_j^n, d_j^n, e_j^n$. The solution to this system supplies an approximate solution to (\ref{etaab}) given in the mesh points $x_j$.

\subsubsection{KdV equation}
In order to demonstrate the construction of the matrices involved we limit at this point our considerations to the first order equation (\ref{kdv1}). It means that we drop temporarily in (\ref{CrABw}) terms of second order,
% in small parameters, 
that is, the terms with $\alpha^2, \alpha\beta, \beta^2$.
Equations with $p$ and $q$ do not apply because $\eta_{4x}$ and  $\eta_{5x}$ do not appear in  (\ref{kdv1}).
This leads to equations 
\begin{eqnarray} \label{matr1}
 \sum_{j=1}^{N} (a^{n\!+\!1}_{j} \!-\! a^{n}_{j}) (\varphi_{j},\psi_{i}) 
\!+\! \tau \frac{1}{2} \sum_{j=1}^{N}(b^{n\!+\!1}_{j} \!+\! b^{n}_{j}) (\varphi_{j},\psi_{i})
 & & \nonumber \\
\!+\! \tau \alpha \frac{3}{16} \sum_{j=1}^{N} \sum_{k=1}^{N}(a^{n\!+\!1}_{j} \!+\! a^{n}_{j}) (a^{n\!+\!1}_{k}  \!+\! a^{n}_{k}) \hspace{10ex}  & & \\  %\hspace{10ex} 
\times \,
(\varphi_{j}'\varphi_{k} \!+\! \varphi_{j}\varphi_{k}',\psi_{i}) \hfill & & \nonumber\\
\!+\!\tau \beta \frac{1}{12} \sum_{j=1}^{N}(c^{n\!+\!1}_{j} \!+\! c^{n}_{j}) (\varphi_{j},\psi_{i}) &=&0 ,  \nonumber \\
\sum_{j=1}^{N} \left[(a^{n\!+\!1}_{j} \!+\! a^{n}_{j}) (\varphi_{j}',\psi_{i}) \!-\! (b^{n\!+\!1}_{j} \!+\! b^{n}_{j}) (\varphi_{j},\psi_{i}) \right] &=&0 ,\nonumber\\
\sum_{j=1}^{N} \left[(b^{n\!+\!1}_{j} \!+\! b^{n}_{j}) (\varphi_{j}',\psi_{i}) \!-\! (c^{n\!+\!1}_{j} \!+\! c^{n}_{j}) (\varphi_{j},\psi_{i}) \right] &=&0 . \nonumber 
\end{eqnarray}
Define 
\begin{equation} \label{defC}\def\arraystretch{1.4}
\begin{array}{ll}
C^{(1)}_{ij}:=(\varphi_{j},\psi_{i}),& C^{(2)}_{ij}:=(\varphi'_{j},\psi_{i}), \\
 C^{(3)}_{ijk}:=(\varphi'_{j}\varphi_{k}+\varphi_{j}\varphi'_{k},\psi_{i}), &
\end{array}
\end{equation}  
where $\varphi'_j=\frac{d\varphi}{dx}(x_j)$. Simple integration shows that 
\begin{equation} \label{C1}\def\arraystretch{1.4}
C^{(1)}_{ij}=\left\{ \begin{array}{rl}
\frac{1}{2} \chi &\quad  \textrm{if} \quad i=j \vee i=j-1 \\
0 & \quad \textrm{otherwise},
\end{array} \right. 
\end{equation}  
\begin{equation} \label{C2}\def\arraystretch{1.4}
C^{(2)}_{ij}=\left\{ \begin{array}{rl}
-1&\quad  \textrm{if} \quad i=j \\
 1&\quad  \textrm{if} \quad i=j-1 \\
0 & \quad \textrm{otherwise}.
\end{array} \right. 
\end{equation}  
 Similarly one obtains
%A little more complicated calculation yields
\begin{equation} \label{C3}\def\arraystretch{1.4}
C^{(3)}_{ijk}= C^{(2)}_{ij}\, \delta_{jk}.
\end{equation}  
The property (\ref{C3}) reduces the double sum in the term with $\tau\alpha\frac{3}{16}$ to the single one of the square of $(a^{n\!+\!1}_{j} \!+\! a^{n}_{j})$.
Insertion of (\ref{C1})--(\ref{C3}) into (\ref{matr1}) gives
\begin{eqnarray} \label{matr2}
 \sum_{j=1}^{N}\left[ (a^{n+1}_{j} \!-\! a^{n}_{j}) C^{(1)}_{ij} 
\!+\! \tau\left( \frac{1}{2} (b^{n+1}_{j} \!+\! b^{n}_{j}) C^{(1)}_{ij} \right. \right. \hspace{4ex}
 \!&\! \!&\!  \\ \left. \left.
+ \alpha \frac{3}{16} (a^{n+1}_{j} \!+\! a^{n}_{j})^2  C^{(2)}_{ij}  %  & & \\
+\beta \frac{1}{12} (c^{n+1}_{j} \!+\! c^{n}_{j}) C^{(2)}_{ij} \right) \right]
 \!  \!& = &\! \!0,  \nonumber \\
\sum_{j=1}^{N} \left[(a^{n+1}_{j} + a^{n}_{j}) C^{(2)}_{ij} - (b^{n+1}_{j} + b^{n}_{j}) C^{(1)}_{ij} \right] &\! \!= &\! \!0, \nonumber\\
\sum_{j=1}^{N} \left[(b^{n+1}_{j} + b^{n}_{j}) C^{(2)}_{ij} - (c^{n+1}_{j} + c^{n}_{j}) C^{(1)}_{ij} \right] &\! \!=\! \!&\! \!0. \nonumber 
\end{eqnarray}

Define the 3$N$-dimensional vector of expansion coefficients 
\begin{equation} \label{vec1}
X^n= \left(\!\begin{array}{c}A^n\\ B^n\\ C^n \end{array}\!\right),
\end{equation} 
where 
\begin{equation} \label{vec2}
A^n= \left(\begin{array}{c} a_1^n\\ a_2^n\\ \vdots \\ a_N^n \end{array} \right)\!,~
B^n= \left(\begin{array}{c} b_1^n\\ b_2^n\\ \vdots \\ b_N^n \end{array} \right)\!,~
C^n= \left(\begin{array}{c} c_1^n\\ c_2^n\\ \vdots \\ c_N^n \end{array} \right)\!.
\end{equation}
In (\ref{matr2}), $A^{n+1}, B^{n+1}, C^{n+1}$ represent the unknown coefficients and $A^{n}, B^{n}, C^{n}$ the known ones. Note that the system (\ref{matr2}) is  nonlinear. The single nonlinear term is quadratic in unknown coefficients. 
For the second order equations (\ref{etaab}) and (\ref{etaabd}) there are more nonlinear terms.

In an abbreviated form the set (\ref{matr2}) can be written as 
\begin{equation} \label{mv1}
F_i(X^{n+1},X^{n}) =  0, \quad i=1,2,\ldots,3N.
\end{equation}
Since this equation is nonlinear we can use the Newton method for each time step. That is, we find $ X^{n+1}$ by iterating the equation
\begin{equation} \label{iter}
(X^{n+1})_{m+1}=(X^{n+1})_{m} + J^{-1}(X^{n+1})_{m}  =  0,
\end{equation}
where $J^{-1}$ is the inverse of the Jacobian of  $F(X^{n+1},X^{n})$   (\ref{mv1}). Choosing $(X^{n+1})_{0} =X^{n}$ we %usually 
obtain the approximate solution to (\ref{mv1}), $(X^{n+1})_m$ in $m=3-5$ iterations with very good precision.
The Jacobian itself is a particular $(3N\times 3N)$ sparse matrix with the following block structure
\begin{equation} \label{J1}
J= \left( \begin{array}{ccc} (A_a) & (A_b) &  (A_c)  \\ (C2) & -(C1) &  (0)   \\ (0) & (C2) & -(C1) \end{array} \right),
\end{equation}
where each block $(\cdot )$ is a two-diagonal sparse $(N\times N)$ matrix. The matrix $A_a$ is given by
\begin{equation} \label{Aa}
A_a\!=\! \left(\! \!\begin{array}{ccccccc} 
a^{1}_{1} & 0 & 0 & \cdots & 0 & a^{1}_{N-1} & a^{1}_{N} \\
a^{2}_{1} &a^{2}_{2} & 0 & \cdots & 0 & 0 & a^{2}_{N} \\
0 & a^{3}_{2} &  a^{3}_{3} & 0 & \cdots & 0 & 0 \\
\vdots & \vdots & \vdots & \ddots & \vdots & \vdots & \vdots \\
0 & 0 & \cdots & a^{N-3}_{N-4} & a^{N-3}_{N-3} & 0 & 0 \\
0 & 0 & \cdots & 0 & a^{n-2}_{N-3} & a^{N-2}_{N-2} & 0 \\
a^{N}_{1} & 0 & \cdots & 0 & 0 & a^{N-1}_{N} & a^{n}_{N} \!
\end{array} \right).
\end{equation}
In (\ref{Aa}) the nonzero elements of $A_a$ are given by 
\begin{equation} \label{Aai}
a^i_j=\frac{\partial\, F_i}{\partial\, a^{n+1}_j}, 
\end{equation}
where $ F$ is given by (\ref{mv1}). The
elements in the upper right and lower left corners result from periodic boundary conditions.
Matrices $A_b$ and $A_c$ have the same structure as  $A_a$, with only elements $a^i_j$ having to be replaced by $b^i_j=\frac{\partial\, F_i}{\partial\, b^{n+1}_j}$ and $c^i_j=\frac{\partial\, F_i}{\partial\, c^{n+1}_j}$, respectively.

Matrices $C1$ and $C2$ are constant. They are defined as
\begin{equation} \label{Ck1}
Ck\!=\! \left(\! \!\begin{array}{ccccc} 
C^{(k)}_{11} & 0  & \cdots &  C^{(k)}_{11} & C^{(k)}_{1N} \\
C^{(k)}_{21} & C^{(k)}_{22}  & \cdots &  0 & C^{(k)}_{2N} \\
%0 & a^{3}_{2} &  a^{3}_{3} & 0 & \cdots & 0 & 0 \\
\vdots & \vdots  & \ddots &  \vdots & \vdots \\
%0 & 0 & \cdots & a^{N-3}_{N-4} & a^{N-3}_{N-3} & 0 & 0 \\
0 & 0 & \cdots & C^{(k)}_{N-1N-1} & 0 \\
C^{(k)}_{N1} & 0 & \cdots  &  C^{(k)}_{N-1N} & C^{(k)}_{NN} \!
\end{array} \right) ,
\end{equation}
where  ~$k=1,2$.

\subsubsection{Extended KdV equation (\ref{etaab})} \label{2nd}

For the second order equation (\ref{etaab}) there are more nonlinear terms. These are terms with $\alpha^2$ and $\alpha\beta$. According to the Petrov--Galerkin scheme we get for the term with  $\alpha^2$
\begin{eqnarray} \label{u3}
& \displaystyle \partial_x &   \left(\eta^{n+\frac{1}{2}}\right)^{3} =  \frac{1}{8} \left(
\partial_x\sum_{j=1}^{N} \left(a_{j}^{n+1} + a_{j}^{n}\right)\varphi_{j}\right)^{3} \nonumber \\
& = &  \frac{1}{8} \partial_x\sum_{j=1}^{N} \sum_{k=1}^{N} \sum_{l=1}^{N} [a_{j}^{n+1}+a_{j}^{n}][a_{k}^{n+1}+a_{k}^{n}] [a_{l}^{n+1}+a_{l}^{n}]   \nonumber\\
&  &  \hspace{20ex} \times \varphi_{j}\varphi_{k}\varphi_{l}  \\
& = &  \frac{1}{8} \sum_{j=1}^{N} \sum_{k=1}^{N} \sum_{l=1}^{N} [a_{j}^{n+1}+a_{j}^{n}][a_{k}^{n+1}+a_{k}^{n}] [a_{l}^{n+1}+a_{l}^{n}]   \nonumber\\
&  &  \hspace{10ex} \times \left(\varphi'_{j}\varphi_{k}\varphi_{l} + \varphi_{j}\varphi'_{k}\varphi_{l} + \varphi_{j}\varphi_{k}\varphi'_{l} \right). \nonumber 
\end{eqnarray}
Denote 
\begin{equation} \label{C4}
C^{(4)}_{ijkl}:= \left(\left[\varphi'_{j}\varphi_{k}\varphi_{l} + \varphi_{j}\varphi'_{k}\varphi_{l} + \varphi_{j}\varphi_{k}\varphi'_{l} \right],\psi_i \right).
\end{equation}
As with $C^{(3)}_{ijk}$ in (\ref{C3}) the following property holds
\begin{equation} \label{C4a}
C^{(4)}_{ijkl} = C^{2}_{ij}\,\delta_{jk}\,\delta_{kl} .
\end{equation}

In a similar way, for terms with $\alpha\beta$ we obtain
\begin{eqnarray} \label{v2}
& \displaystyle \partial_x &   \left(v^{n+\frac{1}{2}}\right)^2   \\  
& = &    \frac{1}{4}
\partial_x\left( \sum_{j=1}^{N} \left(b_{j}^{n+1} + b_{j}^{n}\right)\varphi_{j} \sum_{k=1}^{N} \left(b_{k}^{n+1} + b_{k}^{n}\right)\varphi_{k} \right) \nonumber \\
& = &  \frac{1}{4}
\sum_{j=1}^{N} \sum_{k=1}^{N}  [b_{j}^{n+1}+a_{j}^{n}][b_{k}^{n+1}+b_{k}^{n}] \left( \varphi'_{j}\varphi_{k} + \varphi_{j}\varphi'_{k} \right)   \nonumber 
\end{eqnarray}
and
\begin{eqnarray} \label{u1u2}
& \displaystyle \partial_x &   \left(\eta^{n+\frac{1}{2}}w^{n+\frac{1}{2}}\right)   \\  
& = &    \frac{1}{4}
\partial_x\left( \sum_{j=1}^{N} \left(a_{j}^{n+1} + a_{j}^{n}\right)\varphi_{j} \sum_{k=1}^{N} \left(a_{k}^{n+1} + a_{k}^{n}\right)\varphi_{k} \right) \nonumber \\
& = &  \frac{1}{4}
\sum_{j=1}^{N} \sum_{k=1}^{N}  [a_{j}^{n+1}+a_{j}^{n}][b_{k}^{n+1}+b_{k}^{n}] \left( \varphi'_{j}\varphi_{k} + \varphi_{j}\varphi'_{k} \right).   \nonumber 
\end{eqnarray}
The scalar products appearing in the terms proportional to $\alpha^2$ and $\alpha\beta$ are already defined: $\left(\left( \varphi'_{j}\varphi_{k} + \varphi_{j}\varphi'_{k} \right),\psi_i \right)=C^{(3)}_{ijk}$.

Due to properties (\ref{C4a}) and (\ref{C3}) triple and double sums reduce to single ones.
With these settings the second order KdV equation (\ref{CrABw}) gives the following system  of equations 
\begin{eqnarray} \label{matr3}
 \sum_{j=1}^{N}  \left\{ (a^{n+1}_{j} - a^{n}_{j}) C^{(1)}_{ij} 
+ \tau \left[ \frac{1}{2} (b^{n+1}_{j} + b^{n}_{j}) C^{(1)}_{ij} \right. \right.
\hspace{2ex}
 & &  \\
+  \left( \alpha \frac{3}{16} (a^{n+1}_{j} + a^{n}_{j})^2  
+ \beta \frac{1}{12} (c^{n+1}_{j} + c^{n}_{j}) \right. & &  \nonumber \\
- \alpha^2 \frac{1}{64}  
(a^{n+1}_{j} + a^{n}_{j})^3   
+ \alpha\beta \frac{13}{192}  (b^{n+1}_{j} + b^{n}_{j})^2 & &\nonumber  \\
+\alpha\beta \frac{5}{96}  (a^{n+1}_{j} + a^{n}_{j}) (c^{n+1}_{j}  + c^{n}_{j})  & &\nonumber  \\  \left.  \left. \left.
+ \beta^2 \frac{19}{720}(e^{n+1}_{j}  + e^{n}_{j}) \right) C^{(2)}_{ij} \right]\right\}
&= & 0, \nonumber \\
\sum_{j=1}^{N} \left[(a^{n+1}_{j} + a^{n}_{j}) C^{(2)}_{ij} - (b^{n+1}_{j} + b^{n}_{j}) C^{(1)}_{ij} \right] &=&0, \nonumber\\
\sum_{j=1}^{N} \left[(b^{n+1}_{j} + b^{n}_{j}) C^{(2)}_{ij} - (c^{n+1}_{j} + c^{n}_{j}) C^{(1)}_{ij} \right] &=&0, \nonumber \\
\sum_{j=1}^{N} \left[(c^{n+1}_{j} + c^{n}_{j}) C^{(2)}_{ij} - (d^{n+1}_{j} + d^{n}_{j}) C^{(1)}_{ij} \right] &=&0, \nonumber\\
\sum_{j=1}^{N} \left[(b^{n+1}_{j} + b^{n}_{j}) C^{(2)}_{ij} - (e^{n+1}_{j} \!+\! e^{n}_{j}) C^{(1)}_{ij} \right] &=&0, \nonumber 
\end{eqnarray}
where $i=1,2,\ldots,N$.

In this case the vector of expansion coefficients $X^n$ is $5N$-dimensional 
\begin{equation} \label{vec3}
X^n= \left(\!\begin{array}{c}A^n\\ B^n\\ C^n \\ D^n\\ E^n\end{array}\!\right),
\end{equation} 
where $A^n$, $B^n$ and  $C^n$ are already defined in (\ref{vec2}) and  
\begin{equation} \label{vec4}
D^n= \left(\begin{array}{c} d_1^n\\ d_2^n\\ \vdots \\ d_N^n \end{array} \right)\!,~
E^n= \left(\begin{array}{c} e_1^n\\ e_2^n\\ \vdots \\ e_N^n \end{array} \right)\!.
\end{equation}
The Jacobian becomes now $(5N\times 5N)$ dimensional. Its structure, however, is similar to  (\ref{J1}), that is
\begin{equation} \label{J2}
J= \left( \begin{array}{ccccc} (A_a) & (A_b) &  (A_c) & (0) & (A_e) \\ (C2) & -(C1) &  (0) & (0) & (0)  \\ (0) & (C2) & -(C1) & (0) & (0) \\
(0) & (0) &(C2) & -(C1) & (0) \\  (0) & (0) & (0) &(C2) & -(C1)  
\\\end{array} \right),
\end{equation}
where the matrices $(A_a), (A_b), (A_c)$ are defined as previously and $(A_e)_{ij}=
\frac{\partial F_i}{\partial e^{n+1}_j}$.  
Now $F_i$ represents %the vector corresponding to 
the set (\ref{matr3}) which contains four nonlinear terms.

\subsubsection{Extended KdV equation with uneven bottom} \label{2ndb}

For the extended KdV with non-flat bottom we have to include into (\ref{matr3}) three additional terms contained in the last line of formula (\ref{etaabd}).
% Writing mesh decompositions of the bottom function ${h(x)}$ and its second derivative $h_{2x}(x)$ as
Expanding the bottom function ${h(x)}$ and its second derivative $h_{2x}(x)$ in the basis $\{\varphi_j(x)\}$
\begin{equation} \label{h02}
h(x)=\sum_{j=1}^{N} H0_j \varphi_j(x), \quad h_{2x}(x)=\sum_{j=1}^{N} H2_j \varphi_j(x)
\end{equation}
we can write the terms mentioned above 
% bottom terms 
in the following form
\begin{eqnarray} \label{heta}
 \!&\! \! \partial_x  \!&\! \left(h\eta^{n+\frac{1}{2}}  \right)  \\
 \!&\! \! =  \!&\! \frac{1}{2}
\sum_{j=1}^{N} \sum_{k=1}^{N}  H0_j\,\left(a_{k}^{n+1}+a_{k}^{n}\right) \left( \varphi'_{j}\varphi_{k} + \varphi_{j}\varphi'_{k} \right),  \nonumber \\
 \!&\! \! \partial_x  \!&\! \left(h_{2x}\eta^{n+\frac{1}{2}}  \right)  \label{h2eta}\\ 
 \!&\! \! =  \!&\! \frac{1}{2}
\sum_{j=1}^{N} \sum_{k=1}^{N}  H2_j\,\left(a_{k}^{n+1}+a_{k}^{n}\right) \left( \varphi'_{j}\varphi_{k} + \varphi_{j}\varphi'_{k} \right).  \nonumber \\
 \!&\! \! \partial_x  \!&\! \left(h\eta_{2x}^{n+\frac{1}{2}}  \right)  \label{heta2}\\ 
 \!&\! \! =  \!&\! \frac{1}{2}
\sum_{j=1}^{N} \sum_{k=1}^{N}  H0_j\,\left(c_{k}^{n+1}+c_{k}^{n}\right) \left( \varphi'_{j}\varphi_{k} + \varphi_{j}\varphi'_{k} \right).  \nonumber 
\end{eqnarray}
Since 
$$\left( \left(\varphi'_{j}\varphi_{k} + \varphi_{j}\varphi'_{k} \right),\psi_i \right)=C^{(3)}(i,j,k)=C^{(2)}(i,j)\,\delta_{jk},$$
terms proportional to $\beta\delta$ can be reduced to single sums like those proportional to $\alpha^2, \alpha\beta$ and $\beta^2$ discussed in previous subsections. Finally %taking the  bottom terms into account 
one obtains (\ref{etaabd}) as a system of coupled nonlinear equations ($i=1,2,\ldots,N$)
\begin{eqnarray} \label{matr4}
 \sum_{j=1}^{N}  \left\{ (a^{n+1}_{j} - a^{n}_{j}) C^{(1)}_{ij} 
+ \tau \left[ \frac{1}{2} (b^{n+1}_{j} + b^{n}_{j}) C^{(1)}_{ij} \right. \right.
\hspace{3ex}
 & &  \\
+  \left( \alpha \frac{3}{16} (a^{n+1}_{j} + a^{n}_{j})^2  
+ \beta \frac{1}{12} (c^{n+1}_{j} + c^{n}_{j}) \right. & &  \nonumber \\
- \alpha^2 \frac{1}{64}  
(a^{n+1}_{j} + a^{n}_{j})^3   
+ \alpha\beta \! \left(\! \frac{13}{192}  (b^{n+1}_{j} + b^{n}_{j})^2 \right. & &\nonumber  \\  \left.
+ \frac{5}{96}  (a^{n+1}_{j} \!+\! a^{n}_{j}) (c^{n+1}_{j} \!+\! c^{n}_{j}) \!\right)\!
+ \beta^2 \frac{19}{720}(e^{n+1}_{j}  + e^{n}_{j}) & &\nonumber  \\
-\frac{1}{4} \delta  H0_j\,\left(a_{k}^{n+1}+a_{k}^{n}\right) +\frac{1}{8} \beta\delta  H2_j\,\left(a_{k}^{n+1}+a_{k}^{n}\right) & &\nonumber  \\
\left.  \left. \left. -\frac{1}{8} \beta\delta  H0_j\,(c^{n+1}_{j} + c^{n}_{j})
\right) C^{(2)}_{ij} \right] \! \right\}
&= & 0, \nonumber \\
\sum_{j=1}^{N} \left[(a^{n+1}_{j} + a^{n}_{j}) C^{(2)}_{ij} - (b^{n+1}_{j} + b^{n}_{j}) C^{(1)}_{ij} \right] &=&0, \nonumber\\
\sum_{j=1}^{N} \left[(b^{n+1}_{j} + b^{n}_{j}) C^{(2)}_{ij} - (c^{n+1}_{j} + c^{n}_{j}) C^{(1)}_{ij} \right] &=&0, \nonumber \\
\sum_{j=1}^{N} \left[(c^{n+1}_{j} + c^{n}_{j}) C^{(2)}_{ij} - (d^{n+1}_{j} + d^{n}_{j}) C^{(1)}_{ij} \right] &=&0, \nonumber\\
\sum_{j=1}^{N} \left[(b^{n+1}_{j} + b^{n}_{j}) C^{(2)}_{ij} - (e^{n+1}_{j} \!+\! e^{n}_{j}) C^{(1)}_{ij} \right] &=&0. \nonumber 
\end{eqnarray}
In this case the structures of the vector $X^n$ and all matrices remain the same as in (\ref{vec3})--(\ref{J2}). However the matrix elements in matrices $A_a$ and $A_c$ are now different to those in the previous subsection \ref{2nd}, due to new terms in (\ref{matr4}).% of bottom origin.
 
\begin{figure}[tbh]  
%\begin{center}
\resizebox{1.01\columnwidth}{!}{\includegraphics{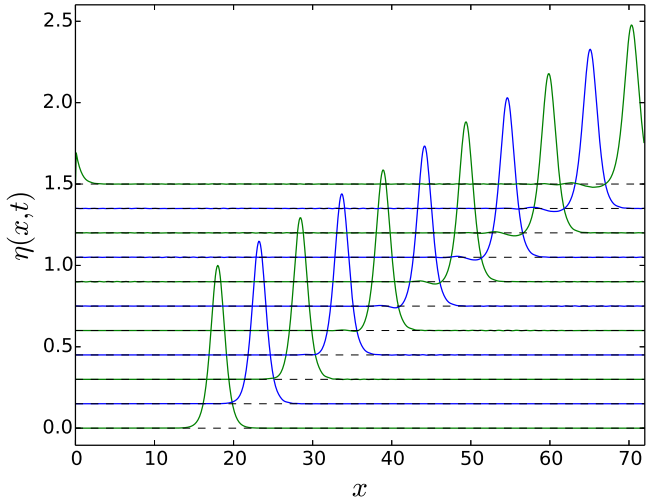}}
%\end{center}
\vspace{-5mm}
\caption{Time evolution of the initial KdV soliton according to the extended KdV equation (\ref{etaab}). Profiles are obtained  by numerical solution of the set of equations (\ref{matr3}). Dashed lines represent the undisturbed fluid surface.} 
 \label{plaskie}
\end{figure}

\section{Numerical simulations} \label{nsym}

It was demonstrated in \cite{DebP} that the method described in the previous section works reasonably well for the KdV equation (\ref{kdvm1}).
Our aim was to apply the finite element method in order to numerically solve the second order equations with a flat bottom (\ref{etaab}) and with an uneven bottom (\ref{etaabd}). There exist two kinds of solutions to KdV equations: soliton (in general, multi-soliton) solutions and periodic solutions called cnoidal waves, see, e.g.\ \cite{Whit,Ding}. In subsections \ref{nsyme} and \ref{unb} we present some examples of numerical simulations for soliton solutions, whereas in the subsection \ref{cno}, some examples for cnoidal solutions.

\subsection{Extended KdV equation (\ref{etaab})} \label{nsyme}

 In Fig.~1 we present several steps of the time evolution of the soliton wave (at $t=0$ it is the KdV soliton) according the the extended KdV equation (\ref{etaab}) and numerical scheme  (\ref{matr3}).
%In Fig.~1 we present the time evolution of the wave with respect to the extended KdV equation (\ref{etaab}) using a numerical scheme according to (\ref{matr3}).  
The mesh size is $N=720$, with a time step $\tau=\chi^2$, %$\tau=0.1$,
and parameters  $\alpha=\beta=0.1$. Plotted are the calculated profiles of the wave $\eta(x,t_k)$ where $t_k=5\cdot k$, \linebreak$k=0,1,...,10$. 
In order to avoid overlaps of profiles at different time instants each subsequent profile is shifted up by 0.15  with respect to the previous one. 
This convention is used in Figs. \ref{gauss} and \ref{2gauss}, as well. Here and in the next figures the dashed lines represent the undisturbed fluid surface.
As the initial condition we chose the standard KdV soliton centered at $x_0=18$. That is, in the applied units,  %\linebreak
$\eta(x,t=0) = \textrm{sech}^2\left[\frac{\sqrt{3}}{2}(x-x_0)\right]$. Note, that since we use scaled variables and  definition (\ref{smallp}) the amplitude of the soliton is equal~1.
In Figs. \ref{gauss}-\ref{well} we use the same initial conditions.

The soliton motion shown in Fig.~1 is in % full 
agreement with the numerical results obtained with the finite difference method in \cite{KRR,KRI}. With parameters $\alpha=\beta=0.1$ the resulting distortion of the KdV soliton due to second order terms in (\ref{etaab}), (\ref{matr3})  is  %hardly visible. created bhind 
in the form of a small amplitude wavetrain created behind the main wave.

\subsection{Uneven bottom  %(\ref{etaabd})
} \label{unb} 

\begin{figure}[bht]
%\begin{center}
\resizebox{1.01\columnwidth}{!}{\includegraphics{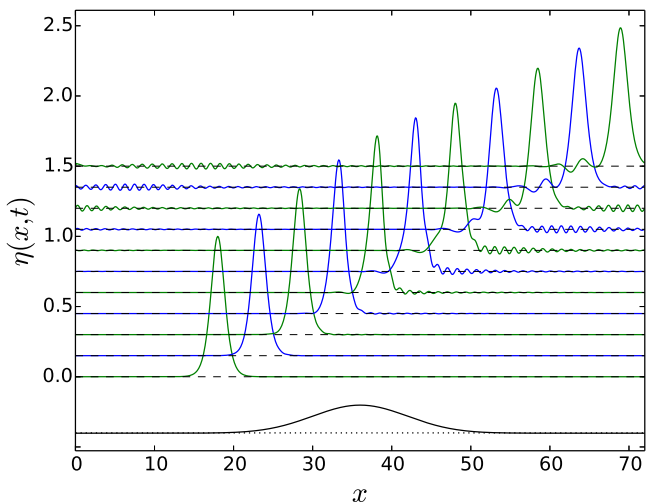}}
%\end{center}
\vspace{-5mm}
\caption{Time evolution of the initial KdV soliton governed by the extended KdV equation (\ref{etaabd}) when the bottom has one hump. Here and in the following figures the dotted line shows the position of (the) undisturbed bottom.} 
 \label{gauss}
\end{figure}

\begin{figure}[tbh]
%\begin{center}
\resizebox{1.01\columnwidth}{!}{\includegraphics{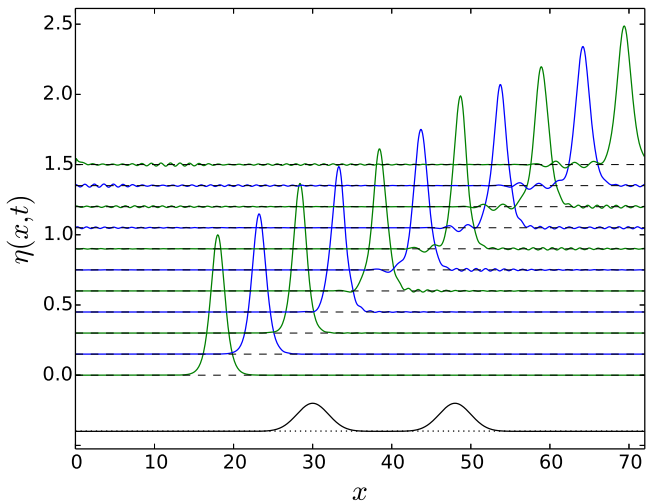}}
%\end{center}
\vspace{-5mm}
\caption{Time evolution of the initial KdV soliton governed by the extended KdV equation (\ref{etaabd}) when the bottom has two narrow humps.} 
 \label{2gauss}
\end{figure}

\begin{figure}[bht]
%\begin{center}
\resizebox{1.01\columnwidth}{!}{\includegraphics{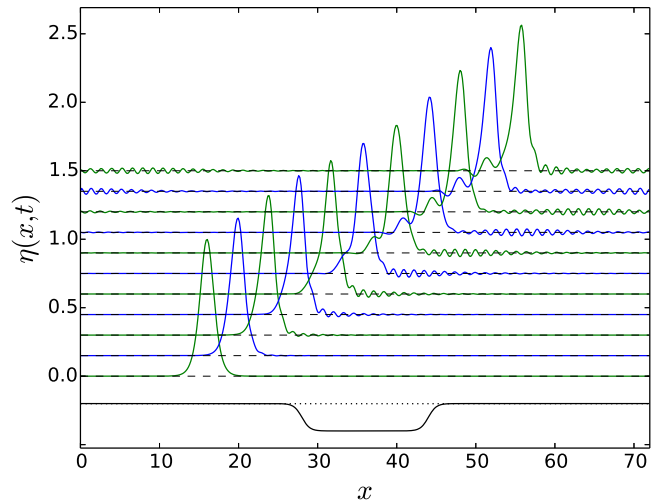}}
%\end{center}
\vspace{-5mm}
\caption{Time evolution of the initial KdV soliton governed by the extended KdV equation (\ref{etaabd}) when the bottom has a well.} 
 \label{well}
\end{figure}

We may question whether the FEM numerical approach to the extended KdV  (\ref{matr4}) is precise enough to reveal the details of soliton distortion caused by a varying bottom. The examples plotted in Figs.~\ref{gauss}-\ref{well} show that it is indeed the case. 
In all the presented calculations the amplitude of the bottom variations is $\delta\!=\!0.2$.
The bottom profile is plotted as a black line below zero on a different scale than the wave profile.

In Fig.~\ref{gauss} the motion of the KdV soliton over a wide bottom hump of Gaussian shape is presented. Here, the bottom function is $h(x)= \delta \exp(-(\frac{x-36}{7})^2)$.
%$e^{-(\frac{x-x_g}{\sigma})^2}$. In the calculations  $\delta=0.2$, $x_g=36$ and $\sigma=20$.
In the scaled variables the undisturbed surface of the water (dashed lines) is at $y=0$. %and the undisturbed bottom is at $-1$. 
The soliton profiles shown in Fig.~\ref{gauss} are almost the same as the profiles obtained with the finite differences method (FDM) used in \cite{KRR,KRI}. 
There are small differences due to smaller precision of our FEM calculations. The FEM allows for the use of larger time steps then FDM. However, in the FEM  the computing time grows rapidly with the increase in the number $N$ of the mesh, since calculation of the inverse of the Jacobian $(5N\times 5N)$ matrices becomes time consuming. \vspace{1mm}

Fig.~\ref{2gauss} displays the motion of the KdV soliton above a double humped Gaussian shaped bottom defined by $h(x)=\delta[\exp(-(\frac{x-30}{6\sqrt{2}})^2)+\exp(-(\frac{x-48}{6\sqrt{2}})^2)$. 
%$h(x)=\delta[\exp(-(x-30)^2/(6\sqrt{2})+\exp(-(x-48)^2/(6\sqrt{2})]$. 
Here  both Gaussians are rather narrow and therefore distortions of the wave shape from the ideal soliton are smaller than those in Fig.~2.  \vspace{0.5ex}

In Fig.~\ref{well} we see the influence of a bottom well with horizontal size extending the soliton's wavelength. The bottom function is chosen as $h(x)=1-\frac{\delta}{2}[\textrm{tanh}(x-28)+\textrm{tanh}(44-x)]$ symmetric with respect to the center of the $x$ interval. 
Fig.\ \ref{well} shows that during the motion above smooth obstacles two effects appear. First, some additional 'waves' of small amplitude, but moving faster than the main solitary wave appear. Second, a wave of smaller amplitude and smaller velocity appears behind the main wave. Both these properties were observed and described in detail in our previous paper \cite{KRI}.

%UWAGA:  w tej chwili  $h(x)$  ma takie samo oznaczenie jak długo\'s\'c odcinka siatki $h$.

\subsection{Motion of cnoidal waves} \label{cno} 

The cnoidal solutions to  KdV equation are expressed by the Jacobi elliptic {\sf cn$^2$} function. 
The explicit formula for cnoidal solutions is,
see, e.g.,\ \cite{Ding}:
\begin{equation} \label{cnsol}
\eta(x,t) = \eta_2 + H \textrm{cn}^2\left(\left. \frac{x-ct}{\Delta} \right\vert m \right),
\end{equation}
where 
\begin{equation} \label{cnsol1}
\eta_2=\frac{H}{m}\left(1-m-\frac{E(m)}{K(m)} \right),\quad \Delta = h\sqrt{\frac{4 m h}{3 H}},
\end{equation}
and
\begin{equation} \label{cnsol2}
c=\sqrt{gh}\left[1+\frac{H}{mh} \left(1-\frac{m}{2} - \frac{3 E(m)}{2 K(m)}\right) \right].
\end{equation}
The solution (\ref{cnsol})-(\ref{cnsol2}) is written in dimensional quantities, where $H$ is the wave height, $h$ is mean water depth, $g$ is the gravitational acceleration and $m$ is an elliptic parameter. $K(m)$ and $E(m)$ are complete elliptic integrals of the first kind and the second kind, respectively. The value of $m\in [0,1]$ governs the shape of the wave. 
 
For $m\to 0$ the cnoidal solution converges to a cosine function.
For $m\to 1$ the cnoidal wave forms peaked crests and flat troughs, such that for $m=1$ the  distance between crests increases to infinity and the cnoidal wave converges to a soliton solution. 
\vspace{0.5ex}

For %our equations 
(\ref{etaabd}) and ({\ref{etaab}}) we have to express the formulas (\ref{cnsol})-(\ref{cnsol2}) in dimensionless variables.

\begin{figure}[bht]
%\begin{center}
\resizebox{1.01\columnwidth}{!}{\includegraphics{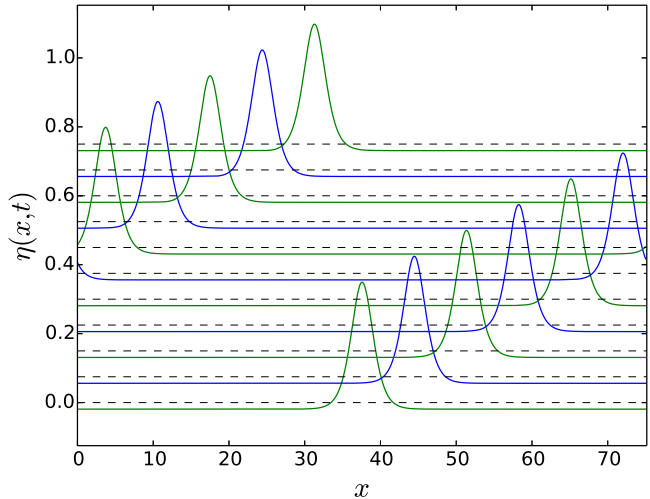}}
%\end{center}
\vspace{-5mm}
\caption{Time evolution of the initial KdV cnoidal wave governed by the extended KdV equation (\ref{etaab}) and numerical scheme (\ref{matr3}).} 
 \label{cn1}
\end{figure}

\begin{figure}[bht]
%\begin{center}
\resizebox{1.01\columnwidth}{!}{\includegraphics{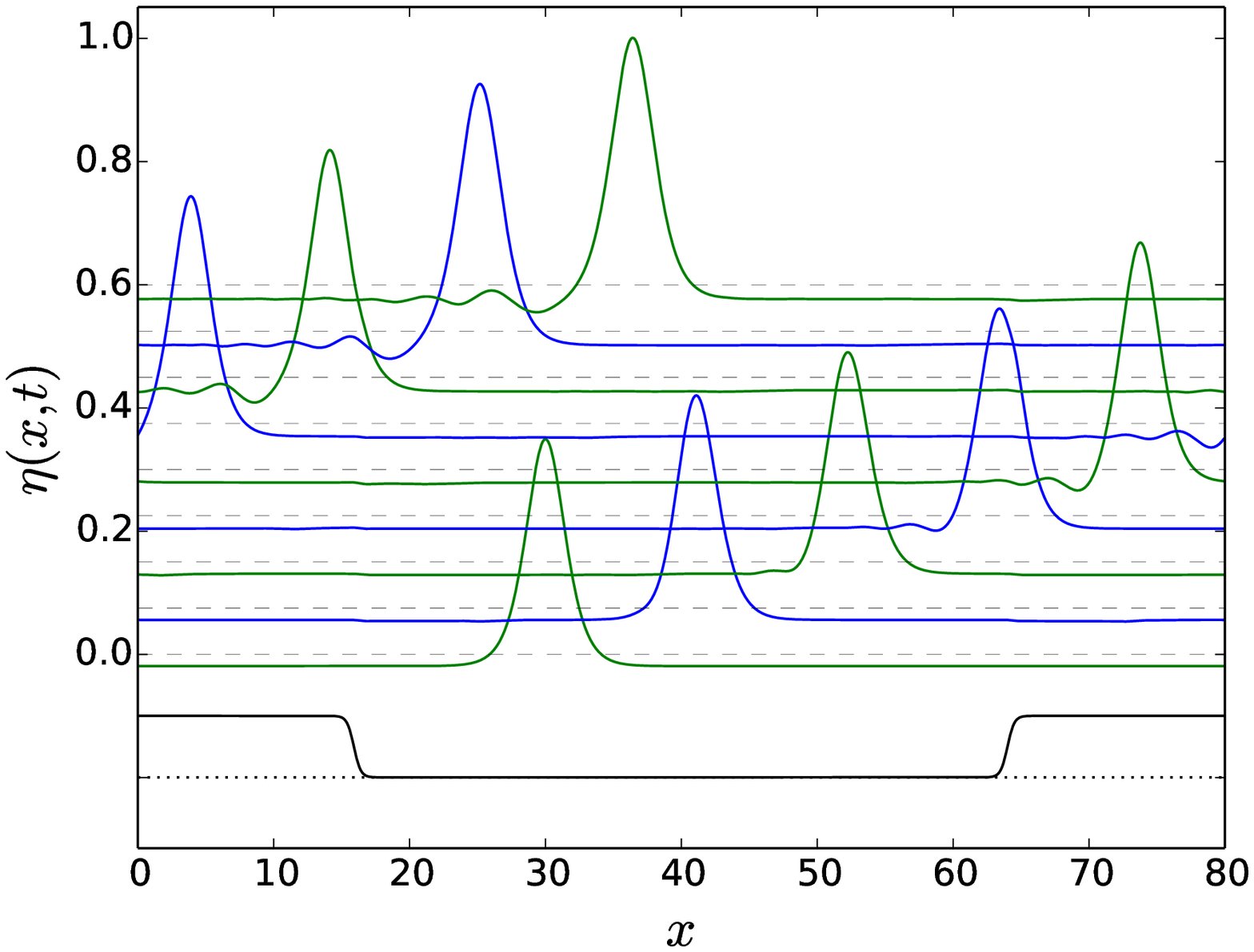}}
%\end{center}
\vspace{-5mm}
\caption{Time evolution of the initial KdV cnoidal wave governed by the extended KdV equation (\ref{etaabd}). The bottom function is here $h(x)=\frac{1}{2}[-\textrm{tanh}(2(x-8.6)-\frac{1}{2})+\textrm{tanh}(2(x-66.5552)-\frac{1}{2})]$.} 
 \label{cn2}
\end{figure}
 
\begin{figure}[tbh]
%\begin{center}
\resizebox{1.01\columnwidth}{!}{\includegraphics{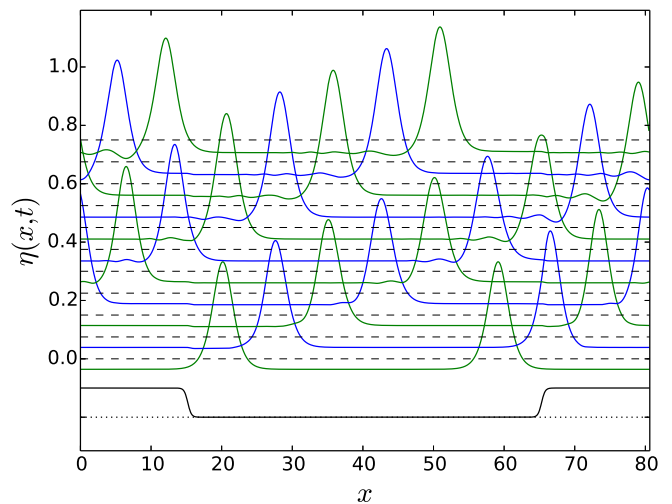}}
%\end{center}
\vspace{-5mm}
\caption{Time evolution of the initial KdV cnoidal wave governed by the extended KdV equation (\ref{etaabd}). The bottom function is here $h(x)=\frac{1}{2}[-\textrm{tanh}(2(x-13.3)-\frac{1}{2})+\textrm{tanh}(2(x-67)-\frac{1}{2})]$.} 
 \label{cn3}
\end{figure}

Fig.\ \ref{cn1} shows the time evolution of the cnoidal wave according to the extended KdV equation (\ref{etaab}), that is, the second order KdV equation with flat bottom. The parameters of the simulation are: $\alpha=\beta=0.14$,~$m\!=\!1\!-\!10^{-16}$. With this value of $m$ the wavelength of the cnoidal wave is equal to $d\approx 75.1552$ %in our 
dimensionless units, and calculations were performed on the interval of that length, $x\in[0,75.1552]$ with a periodic boundary condition. The mesh size was taken as $N=752$. 
The initial position of the wave peak was chosen at the center of chosen interval, that is $x_0=37.5776$. The explicit form of the initial condition in this case was $\eta(x,t=0)= -0.0189862 + 0.368486 \,\textrm{cn}^2\left(\left. \frac{x-x_0}{1.90221} \right\vert m \right) $. Profiles of the wave are plotted at time instants $t_k=10\cdot k$, where $k=0,1,...,8$. % with green lines for even $k$ and blue lines for odd $k$. 
Since the amplitudes of cnoidal waves are smaller than 1, the vertical shift for  the sequential profiles in Figs.\ \ref{cn1}-\ref{cn3} is chosen to be 0.075.

In Fig.\ \ref{cn2} we display the initially cnoidal wave moving over an extended, almost flat hump. In this simulation the value of parameters $\alpha,\beta,m$ and $x$ interval are the same as in the previous figure. Since we consider here the motion over an uneven bottom defined by the function  $h(x)=\frac{1}{2}[-\textrm{tanh}(2(x-8.6)-\frac{1}{2})+\textrm{tanh}(2(x-66.5552)-\frac{1}{2})]$ the evolution was calculated according to equation (\ref{etaabd}) and numerical scheme (\ref{matr4}). Profiles of the wave are plotted at time instants $t_k=10\cdot k$, where $k=0,1,...,8$. % with green lines for even $k$ and blue lines for odd $k$. 
Fig.\ \ref{cn2} shows that  during the %peak
 wave motion over the obstacle a kind of slower wave with smaller amplitude is created following the main peak.

In Fig.\ \ref{cn3} we present the initially cnoidal wave moving over an extended, almost flat hump. In this simulation  $m\!=\!1\!-\!10^{-8}$. The intial condition is given by 
$\eta(x,t=0)= - 0.0359497 + 0.368486 \, \textrm{cn}^2(\frac{x-x_0}{1.90221}|m)$ with $x_0=20.1571$.
Because $m$ is smaller than in the previous cases, the wavelength $d$ of the cnoidal wave is also smaller,  $d\approx 40.3241$.
Calculations were made on the interval $x\in [0,2d] $ with $N=807$.  Profiles of the wave are plotted at time instants $t_k=10\cdot k$, where $k=0,1,...,8$. % with green lines for even $k$ and blue lines for odd $k$. 
%Fig.\ \ref{cn3} shows that during the motion of the wave peak over the obstacle a kind of slower wave with smaller amplitude is created following the main peak.
Fig.\ \ref{cn3} shows qualitatively similar features to those in Fig.\ \ref{cn2}.

\begin{figure}[bht]
%\begin{center}
\resizebox{1.0\columnwidth}{!}{\includegraphics{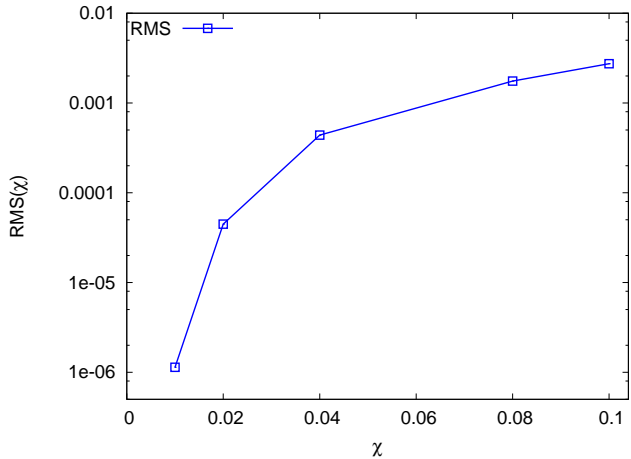}}
%\end{center}
\vspace{-5mm}
\caption{Precision of numerical calculations for KdV equation in the fixed frame as a function of mesh size.} 
 \label{ff8}
\end{figure}

\subsection{Precision of numerical calculations}

The KdV equation (\ref{kdvm}) or  (\ref{kdvm1}) is unique since it possesses an infinite number of invariants, see, e.g.,\ \cite{MGK,DrJ}.  
The lowest invariant, \linebreak
$I_1\!=\!\int_{-\infty}^{+\infty} \eta dx$,  
represents the conservation law for the mass (volume) of the liquid. The second,  $I_2\!=\!\int_{-\infty}^{+\infty} \eta^2 dx$, is related to
 momentum conservation, and the third, \linebreak
$I_3\!=\!\int_{-\infty}^{+\infty} (\eta^3-\frac{1}{3}\eta_x^2) dx$, is related to energy conservation.
However, as pointed by \cite{AbSe,AlKa,KRI2}, the relations between  $I_2$ and momentum and  $I_3$ and energy are more complex.

Approximate conservation of these invariants serves often as a test
of the precision of numerical simulations.
However, this is not the case for the second order KdV type equations (\ref{etaabd}) and  (\ref{etaab}). It was noted in \cite{KRI2} that $I_1$ is an
invariant of equations 
% conserved by solutions to 
(\ref{etaabd}) and  (\ref{etaab}) but  $I_2$ and $I_3$ are not invariants.  Therefore, only $I_1$ can be used as a test for the precision of numerical calculations of waves moving according to the second order extended KdV equations. 
In all the presented calculations the precision of the numerical values of $I_1$ was consistently high (the values  $%\displaystyle 
\frac{I_1(t)-I_1(0)}{I_1(0)}\le 10^{-6}$).  

%%%%%%%%%%%%%%%%%%%%%%%%%%%%%%%%%%%%%%%%%

Wave motion according to KdV and extended (second order) KdV equations is usually
calculated in the reference frame moving with the natural velocity $c=1$ in scaled dimensionless variables (in original variables $c=\sqrt{gh}$). The KdV and extended KdV equations for 
a moving reference frame are obtained by the transformation $\hat{x}=(x-t),~~ \hat{t}=t$ which removes the term $\eta_x$ from the equation (\ref{etaab}). Then the soliton velocity in the fixed frame is proportional to $1+\frac{\alpha}{2}$ whereas in the moving frame it is proportional to $\frac{\alpha}{2}$. Therefore, for value of $\alpha=0.1$ the distance covered by a soliton in the moving frame is $\frac{\alpha}{2}/(1+\frac{\alpha}{2})=\frac{1}{21}$ times shorter than the distance covered in the fixed frame for the same duration. Then, with the same number of the mesh points $N$ the mesh size $\chi$ can be more than 20 times smaller assuring a much higher precision of calculation in the moving frame at the same operational cost. % (calculation time with the same method). 
For instance \cite{DebP} obtained a good precision for motion of KdV soliton with the FEM method using $N=200$,  $\chi=0.01$ and time step $\tau=\chi$ on the interval $x\in[0,2]$.

Precision of FEM method in the fixed frame can be tested by calculation of a root mean square  (RMS) of deviations of wave profile obtained numerically from those obtained from the analytic solution. Denote by 
$\eta_i^{anal}(t)$ and $\eta_i^{num}(t)$ the values of the solutions at given mesh point $i$ an time instant $t$, analytic and numerical, respectively. Then the RMS is expressed as 
\begin{equation}\label{var}
\textrm{RMS}(\chi,t) = \left(\frac{1}{N}\sum_{i=1}^{N} (\eta_i^{anal}(t)-\eta_i^{num}(t))^2\right)^{1/2}
\end{equation}

 We checked our implementation of the FEM on the interval $x\in[0,20]$ using several different sizes $\chi$ of the mesh and several time values.  
% In order to increase  precision we applied time steps $\tau=\chi^2$.
Fig.\ \ref{ff8} displays the RMS (\ref{var})  values for $t=10$. It shows that deviations from
analytic solution decrease substantialy with decreasing~$\chi$. Small $\chi$  assures a very high precision in numerical simulations, however, at the expense of large computation time.
Another tests (not shown here) in which $\chi$ was fixed and RMS was calculated as a function of time showed that for $\tau=\chi^2$ RMS increases with time linerly and very slowly.
 
When the bottom is not flat simulations {\em have to be done in the fixed reference frame}. For our purposes we needed to choose the $x$ intervals of the order of 70 or 80. Even for $\chi=0.1$ the size of Jacobian matrices (\ref{J2}) reaches (4000$\times$4000) and its inversion is time consuming. In a compromise between numerical precision and reasonable computing times we made our simulations with $\chi=0.1$. This choice resulted in about one week of computing time for a single run on the cluster. 
In spite of the insufficient precision the results presented in Figs. 1-7 reproduce details of evolution known from our previous studies,  obtained with the finite difference method. These details, resulting from second order terms in extended KdV (\ref{etaab}), are seen in Fig.~\ref{plaskie} as a wavetrain of small amplitude created behind the main one (compare with Fig.~2 in \cite{KRI}). A similar wavetrain behind the main one was observed in numerical simulations by \cite{MS96}, see e.g.\ Fig.~2 therein.
For waves moving with presence of bottom obstacle these secondary waves behind the main one are amplified by interaction with the bottom and new faster secondary waves appear (see, 
e.g., Figs.\ 2-4). These effects were already observed by us, see Figs.\ 6 and 7  in  \cite{KRI}.

\vspace{3mm}
\noindent {\bf Conclusions}\\[2mm]
The main conclusions of our study can be summarized as follows.
\begin{itemize}
%\item A weak formulation of the finite element method (FEM) for nonlinear equations of second order KdV type (both with an even and uneven bottom) can be effectively used for numerical calculations of the time evolution of both soliton and cnoidal waves.

\item A weak formulation of the finite element method (FEM) for extended KdV equation 
(\ref{etaab}) can be effectively used for numerical calculations of the time evolution of both soliton and cnoidal waves when calculations are done in a moving frame.

\item  Since numerical calculations for equation (\ref{etaabd}) have to be performed in a  fixed frame, the presented FEM method is not as effective as the FDM method used by us in previous papers because the computer time necessary for obtaining sufficiently high  precision becomes impractical. On the other hand, the presented results (though not as precise as FDM ones) exhibit all secondary structures generated by higher order terms of the equations.

\item First tests of numerical solutions to second order KdV type equations with a stochastic term seem to be very promising \cite{KRSB2}. 

\end{itemize}

%\begin{acknowledgment}
The authors would like to thank anonymous referees for several helpful suggestions and remarks that affected the article content.
%\end{acknowledgment}

\end{document}